\documentstyle[version2,preprint,aps]{revtex}                         

\tightenlines
\begin{document}
\draft

\preprint{SLAC--PUB--6146}
\medskip
\preprint{WIS--93/33/May--PH}
\medskip
\preprint{May 1993}
\medskip
\preprint{T/E}

\begin{title}
The Subleading Isgur-Wise Form Factor $\xi_3(v\cdot v')$\\
and its Implications for the Decays $\bar B\to D^{(*)}\ell\,\bar\nu$
\end{title}

\author{Zoltan Ligeti and Yosef Nir}
\begin{instit}
Weizmann Institute of Science\\
Physics Department, Rehovot 76100, Israel
\end{instit}

\author{Matthias Neubert}
\begin{instit}
Stanford Linear Accelerator Center\\
Stanford University, Stanford, California 94309
\end{instit}


\begin{abstract}
We calculate, in the framework of QCD sum rules and to next-to-leading
order in perturbation theory, the universal function $\xi_3(v\cdot v')$
which appears at order $1/m_Q$ in the heavy quark expansion of meson
weak decay form factors. We find that radiative corrections of order
$\alpha_s$ are very important. Over the kinematic range accessible in
semileptonic decays, $\xi_3(v\cdot v')$ is proportional to the
leading-order Isgur-Wise function $\xi(v\cdot v')$ to very good
accuracy. Taking into account all sources of uncertainty, we estimate
$\xi_3/\xi=(0.6\pm 0.2)$. This reduces the theoretical uncertainty in
the extraction of $|\,V_{cb}|$ from $\bar B\to D\,\ell\, \bar\nu$
transitions. A measurement of the form factor ratio $A_2/A_1$ in $\bar
B\to D^*\ell\,\bar\nu$ decays can be used to test our prediction.
\end{abstract}


\centerline{(submitted to Physical Review D)}
\newpage
\narrowtext

\section{Introduction}

In the limit where the charm and bottom quarks are considered
infinitely heavy, their strong interactions with light quarks and
gluons acquire additional symmetries \cite{Shur,Nuss,Volo,Isgu}. In
particular, the weak decay form factors describing the semileptonic
transitions $\bar B\to D\,\ell\,\bar\nu$ and $\bar B\to
D^*\ell\,\bar\nu$ become related to a single universal function. The
normalization of this so-called Isgur-Wise form factor is known in the
zero recoil limit, where the initial and final meson have the same
velocity. This allows a model-independent determination of the
Cabibbo-Kobayashi-Maskawa matrix element $|\,V_{cb}|$ \cite{Volo,Vcb},
up to corrections arising from the fact that $m_c$ and $m_b$ are, after
all, not infinitely heavy. In this work, we investigate a particular
type of such corrections, which is important to the extraction of
$|\,V_{cb}|$ from $\bar B\to D\,\ell\,\bar\nu$ transitions.

The heavy quark effective theory (HQET) provides a convenient framework
to analyze the weak decays of hadrons containing a heavy quark
\cite{Eich,Grin,Geor,Mann,Falk,Luke,FGL}. It provides a systematic
expansion of hadronic matrix elements in powers of $1/m_Q$. The
coefficients in this expansion are universal functions of the
velocities $v$ and $v'$ of the initial and final hadrons ($v^2=v'^2=1$,
$v\cdot v'\ge 1$), but they do not depend on the flavor or spin of the
heavy quarks. These form factors originate from long-distance hadronic
dynamics, so they can only be investigated using some nonperturbative
approach to QCD. One such method is provided by QCD sum rules
\cite{SVZ}, which have recently been widely used to calculate hadronic
matrix elements in HQET
\cite{Buch,Rady,Broa,Baga,Neu1,Neu2,Blok,chi23,Neu3}. At leading order
in the heavy quark expansion, a single Isgur-Wise function $\xi(v\cdot
v',\mu)$ is required to parameterize the current-induced transitions
$M(v)\to M'(v')$, where $M$ and $M'$ are pseudoscalar or vector mesons
containing a single heavy quark \cite{Isgu}. This is conveniently
expressed by the compact trace formula \cite{Falk,Bjor}
\begin{equation}\label{xidef}
   \langle M'(v')|\,\bar h'\,\Gamma\,h\,|M(v)\rangle
   = - \xi(v\cdot v',\mu)\,{\rm Tr}\big\{\,\overline{\cal{M}}'(v')\,
   \Gamma\,{\cal{M}}(v)\,\big\} \,,
\end{equation}
where $\Gamma$ is an arbitrary Dirac matrix, and $h$ and $h'$ denote
the velocity-dependent effective fields in HQET which represent heavy
quarks $Q$ and $Q'$ moving at the hadron's velocities $v$ and $v'$. The
heavy mesons are represented by covariant tensor wave functions
\begin{equation}
   {\cal{M}}(v) = \sqrt{m_M}\,{(1+\rlap/v)\over 2}\,
   \cases{ -\gamma_5 &; pseudoscalar meson, \cr
           \rlap/\epsilon &; vector meson, \cr}
\end{equation}
which have the correct transformation properties under Lorentz boosts
and heavy quark spin rotations. Here $m_M$ denotes the physical meson
mass, and $\epsilon$ is the polarization vector of the vector meson.
Current conservation implies that the Isgur-Wise function is normalized
at zero recoil: $\xi(1,\mu)=1$. Except at this point, the universal
form factor depends on a subtraction scale $\mu$, since the
velocity-changing currents in the effective theory have to be
renormalized. The $\mu$-dependence of the Isgur-Wise function cancels
against that of the Wilson coefficients which appear in the matching of
currents of the full theory onto currents of the effective theory
\cite{Falk,QCD}.

At order $1/m_Q$, matrix elements receive contributions from higher
dimension operators in the effective Lagrangian and in the effective
currents \cite{Luke}. The former give rise to three new universal
functions, usually denoted by $\chi_i(v\cdot v',\mu)$ for $i=1,2,3$. In
the framework of QCD sum rules, we have investigated these form factors
in Refs.~\cite{Neu2,chi23}. Here we shall focus on the second type of
corrections, which come from operators of dimension four in the
short-distance expansion of the weak currents in HQET. In the case of
the vector or axial vector currents, there are fourteen independent
operators of this type. Their Wilson coefficients have recently been
calculated to next-to-leading order in renormalization-group improved
perturbation theory \cite{RPI}. On dimensional grounds, any dimension
four current operator must contain a covariant derivative acting on one
of the heavy quark fields. Thus, these operators have the generic form
$\bar h'\,\Gamma\,i D_\alpha h$ or $(i D_\alpha \bar h')\,\Gamma\,h$,
where $\Gamma$ is again an arbitrary Dirac matrix. In analogy to
(\ref{xidef}), the corresponding matrix elements can be parameterized
by a tensor form factor $\xi_\alpha$ defined by \cite{Luke}
\begin{equation}\label{xial}
   \langle M'(v')|\,\bar h'\,\Gamma\,i D_\alpha h\,|M(v)\rangle
   = - \bar\Lambda\,{\rm Tr}\big\{\,\xi_\alpha(v,v',\mu)\,
   \overline{\cal{M}}'(v')\,\Gamma\,{\cal{M}}(v)\,\big\} \,,
\end{equation}
where $\bar\Lambda=m_M-m_Q=m_{M'}-m_{Q'}$ denotes the mass difference
between a heavy meson and the heavy quark that it contains. This
parameter sets the canonical scale for power corrections in HQET.
Matrix elements with a derivative acting on $\bar h'$ are related to
(\ref{xial}) by Dirac conjugation. The most general decomposition of
the form factor can be written
\begin{equation}\label{xialdef}
   \xi_\alpha(v,v',\mu) = \xi_+(v\cdot v',\mu)\,(v+v')_\alpha
   + \xi_-(v\cdot v',\mu)\,(v-v')_\alpha
   - \xi_3(v\cdot v',\mu)\,\gamma_\alpha \,.
\end{equation}
Due to $T$-invariance of the strong interactions the coefficient
functions $\xi_i(v\cdot v',\mu)$ are real. Furthermore, the equation of
motion of HQET can be used to derive the constraints \cite{Luke}
\begin{eqnarray}\label{xirel}
   \xi_-(v\cdot v',\mu) &=& {1\over 2}\,\xi(v\cdot v',\mu) \,,
    \nonumber\\
   \phantom{ \Bigg[ }
   \xi_+(v\cdot v',\mu) &=& {1\over 2}\,
    {v\cdot v'-1\over v\cdot v'+1}\,\xi(v\cdot v',\mu)
    - {1\over v\cdot v'+1}\,\xi_3(v\cdot v',\mu) \,,
\end{eqnarray}
These relations show that only one of the coefficient functions, say
$\xi_3(v\cdot v',\mu)$, is independent. They also suggest a close
relation between the subleading functions $\xi_i(v\cdot v',\mu)$ and
the leading-order Isgur-Wise function $\xi(v\cdot v',\mu)$. The origin
of this relation is the invariance of the effective theory under
reparameterizations of the heavy quark momentum \cite{RPI,LuMa}. In
fact, one can show that the $\mu$-dependence of the functions
$\xi_i(v\cdot v',\mu)$ is the same as the $\mu$-dependence of the
Isgur-Wise function. This leads us to introduce a new,
renormalization-group invariant function $\eta(v\cdot v')$ by
\begin{equation}\label{etadef}
   \eta(v\cdot v') \equiv {\xi_3(v\cdot v',\mu)\over\xi(v\cdot v',\mu)}
   \,.
\end{equation}
We expect that $\eta(v\cdot v')$ is a slowly varying function of order
unity. By means of (\ref{xirel}) and (\ref{etadef}), meson matrix
elements of the dimension four operators in the short-distance
expansion of the currents can be parameterized in terms of the product
$\bar\Lambda\,\xi(v\cdot v',\mu)$ and a single new function
$\eta(v\cdot v')$, which is independent of the renormalization scale.

Knowledge of the function $\eta(v\cdot v')$ becomes important when one
wants to extract $|\,V_{cb}|$ from a measurement of the differential
decay rate ${\rm d}\Gamma/{\rm d}(v\cdot v')$ for $\bar B\to D\,\ell\,
\bar\nu$ transitions near zero recoil. Because of the known
normalization of the Isgur-Wise function at $v\cdot v'=1$, hadronic
uncertainties affect such a measurement only at the level of power
corrections of order $1/m_c$ and $1/m_b$. Note that whereas the $\bar
B\to D^*\ell\,\bar\nu$ decay rate is protected against first-order
power corrections by Luke's theorem \cite{Luke}, the rate for $\bar
B\to D\,\ell\,\bar\nu$ is not, due to its helicity suppression at zero
recoil \cite{Riec}. However, one can show that the symmetry-breaking
corrections are parametrically suppressed by the ``Voloshin-Shifman
factor'' \cite{Volo}
\begin{equation}
   S = \bigg({m_B-m_D\over m_B+m_D}\bigg)^2 \approx 0.23 \,.
\end{equation}
For this reason one may hope that the theoretical uncertainty in
extracting $|\,V_{cb}|$ from these transitions is not much worse than
in the case of $\bar B\to D^*\ell\,\bar\nu$ decays \cite{Neu2}. An
extrapolation of the spectrum to zero recoil gives
\begin{equation}\label{BDrate}
   \lim_{v\cdot v'\to 1} {1\over\big[(v\cdot v')^2-1\big]^{3/2}}\,
   {{\rm d}\Gamma(\bar B\to D\,\ell\,\bar\nu)\over{\rm d}(v\cdot v')}
   = {G_F^2\,|\,V_{cb}|^2\over 48\pi^3}\,(m_B+m_D)^2\,m_D^3\,
   \Big[ 1 + S\cdot K \Big]^2 \,,
\end{equation}
where
\begin{equation}\label{Kfac}
   K = \delta_1 + \bigg( {\bar\Lambda\over 2 m_c}
   + {\bar\Lambda\over 2 m_b} \bigg) \Big[ (1+\delta_1)
   - 2\,(1+\delta_2)\,\eta(1) \Big] \,.
\end{equation}
Here $\delta_i$ are perturbative QCD corrections arising from finite
renormalizations of the currents in the intermediate region
$m_b>\mu>m_c$. Numerically, one finds $\delta_1\approx 6\%$ and
$\delta_2\approx 9\%$ \cite{QCD,RPI}. The only unknown quantity (except
the heavy quark masses) in this equation is the value of the function
$\eta(v\cdot v')$ at zero recoil.

It has been pointed out in Ref.~\cite{Neu2} that certain ratios of the
hadronic form factors describing $\bar B\to D^*\ell\,\bar\nu$
transitions are very sensitive to symmetry-breaking corrections to the
heavy quark limit. Their measurement can be used to test model
calculations of the subleading universal form factors. Consider the
quantity
\begin{equation}
   R_2 = \bigg[ 1 - {q^2\over(m_B + m_{D^*})^2} \bigg]\,
   {A_2(q^2)\over A_1(q^2)} \,,
\end{equation}
where $A_1(q^2)$ and $A_2(q^2)$ are axial vector form factors in the
notation of Ref.~\cite{BSW}. Introducing the variable $v\cdot v'$
instead of $q^2$, and performing a $1/m_Q$ expansion of the hadronic
form factors, one obtains \cite{Neu2}
\begin{equation}\label{R2def}
   R_2 = 1 - {\bar\Lambda\over v\cdot v'+1}\,\bigg( {1\over m_c}
    + {3\over m_b} \bigg)\,\eta(v\cdot v') + \ldots \,.
\end{equation}
The ellipsis represents a small ``hyperfine correction'' proportional
to the function $\chi_2(v\cdot v',\mu)$, and higher-order power
corrections of order $1/m_Q^2$. The perturbative corrections to $R_2$
turn out to be completely negligible. Unless the function $\eta(v\cdot
v')$ were highly suppressed, the dominant symmetry-breaking correction
to $R_2$ is the one shown in (\ref{R2def}). A measurement of this ratio
would therefore provide valuable information about the form factor
$\eta(v\cdot v')$.

In Ref.~\cite{Neu2}, the function $\xi_3(v\cdot v',\mu)$ has been
analyzed using the QCD sum rule approach and adopting the standard
approximations, in which one neglects radiative corrections. Such an
approach leads to the parameter-free prediction that
$\xi_3(1,\mu)=\eta(1)=1/3$. In the context of QCD sum rules,
corrections to this simple result can only come from radiative
corrections or higher dimension condensates. It was recently found that
such corrections can be quite significant, however. For the subleading
Isgur-Wise functions $\chi_2(v\cdot v',\mu)$ and $\chi_3(v\cdot
v',\mu)$, the terms of order $\alpha_s$ even give the dominant
contributions to the sum rules \cite{chi23}. In the case of
$\xi_3(v\cdot v',\mu)$ this may be expected to be even more so, as due
to the peculiar trace structure associated with this form factor in
(\ref{xialdef}) there is no quark condensate contribution at tree
level. Therefore, it is important to refine the sum rule analysis of
Ref.~\cite{Neu2} by including radiative corrections. This requires, in
particular, the calculation of the two-loop corrections to the triangle
quark-loop diagram. The techniques to handle the corresponding two-loop
integrals have been developed in Ref.~\cite{Neu3}.

\section{Derivation of the Sum Rule}

The QCD sum rule analysis of $\xi_3(v\cdot v',\mu)$ proceeds in
complete analogy to that of the Isgur-Wise function. For a detailed
description of the procedure the reader is referred to
Refs.~\cite{Neu2,Neu3}. Here we shall only very briefly sketch the main
steps. One considers, in the effective theory, the three-current
correlator
\begin{eqnarray}\label{correl}
   &&\int\!{\rm d}x\,{\rm d}x'\,e^{i(k'\cdot x'-k\cdot x)}\,
    \langle\,0\,|\,T\,\Big\{
    \big[ \bar q\,\overline{\Gamma}_{M'} h' \big]_{x'},
    \big[ \bar h'\,\Gamma\,i D_\alpha h \big]_0,
    \big[ \bar h\,\Gamma_M\,q \big]_x \Big\}\,|\,0\,\rangle \nonumber\\
   &&\qquad = {\rm Tr}\Big\{\,\Xi_\alpha(v,v',k,k')\,
    \overline{\Gamma}_{M'} {1+\rlap/v'\over 2}\,\Gamma\,
    {1+\rlap/v\over 2}\,\Gamma_M\,\Big\} \,,
\end{eqnarray}
where $v$ and $v'$ are the velocities of the heavy quarks, and $k$ and
$k'$ are the external off-shell momenta injected into the three-point
function. Depending on the choice $\Gamma_M=-\gamma_5$ or
$\Gamma_M=\gamma_\mu-v_\mu$, the heavy-light currents interpolate
pseudoscalar or vector mesons, respectively. The Dirac structure of the
correlator, as shown in the second line, is a consequence of heavy
quark symmetry, as reflected in the Feynman rules of HQET. The quantity
$\Xi_\alpha$ obeys a decomposition analogous to (\ref{xialdef}), with
coefficient functions $\Xi_\pm$ and $\Xi_3$ that are analytic in the
``off-shell energies'' $\omega=2 v\cdot k$ and $\omega'=2 v'\cdot k'$,
with discontinuities for positive values of these variables. These
functions also depend on the velocity transfer $y=v\cdot v'$. From now
on we shall focus on the coefficient $\Xi_3(\omega,\omega',y)$, which
is used to construct the sum rule for the subleading form factor
$\xi_3(y,\mu)$.

The idea of QCD sum rules is to relate a theoretical approximation to
the operator product expansion of $\Xi_3$ to a hadronic representation
of the correlator in terms of physical intermediate states. The
lowest-lying states are the ground-state mesons $M$ and $M'$ associated
with the heavy-light currents. They lead to a double pole located at
$\omega=\omega'= 2\bar\Lambda$. The residue is proportional to the
function $\xi_3(y,\mu)$. One finds \cite{Neu2}
\begin{equation}\label{pole}
   \Xi_3^{\rm pole}(\omega,\omega',y) = {\bar\Lambda\,\xi_3(y,\mu)\,
   F^2(\mu) \over (\omega-2\bar\Lambda+i\epsilon)
   (\omega'-2\bar\Lambda+i\epsilon)} \,,
\end{equation}
where $F$ corresponds to the scaled meson decay constant ($F\sim
f_M\sqrt{m_M}$). Both $F$ and $\xi_3$ are defined in terms of matrix
elements in the effective theory and are scale-dependent quantities. In
the deep Euclidean region, the correlator can be calculated
perturbatively because of asymptotic freedom. The idea of Shifman,
Vainshtein, and Zakharov was that, at the transition from the
perturbative to the nonperturbative regime, confinement effects can be
accounted for by including the leading power corrections in an operator
product expansion. They are proportional to vacuum expectation values
of local quark-gluon operators, the so-called condensates \cite{SVZ}.
Following the standard procedure, we write the theoretical expression
for $\Xi_3$ as a double dispersion integral and perform a Borel
transformation in $\omega$ and $\omega'$. This yields an exponential
damping factor in the dispersion integral and eliminates possible
subtraction polynomials. Because of the flavor symmetry of HQET, it is
natural to set the associated Borel parameters equal: $\tau=\tau'\equiv
2 T$. Following Ref.~\cite{Blok}, we then introduce new variables
$\omega_+=\case{1}/{2}(\omega+\omega')$ and $\omega_-=\omega-\omega'$,
perform the integral over $\omega_-$, and employ quark-hadron duality
to equate the remaining integral over $\omega_+$ up to a threshold
$\omega_0$ to the Borel transform of the pole contribution in
(\ref{pole}). This yields the Borel sum rule
\begin{equation}\label{sumrul}
   \bar\Lambda\,\xi_3(y,\mu)\,F^2(\mu)\,e^{-2\bar\Lambda/T}
   = \int\limits_0^{\omega_0}\!{\rm d}\omega_+\,e^{-\omega_+/T}\,
   \widetilde{\rho}(\omega_+,y) \equiv K(T,\omega_0,y) \,.
\end{equation}
The effective spectral density $\widetilde{\rho}$ arises after
integration of the double spectral density over $\omega_-$. For
practical purposes it is useful to notice that the
$\omega_+$-dependence of each term in $\widetilde{\rho}(\omega_+,y)$ is
known on dimensional grounds \cite{chi23,Neu3}. It thus suffices to
calculate the Borel transform of $\Xi_3$, corresponding to the limit
$\omega_0\to\infty$ in (\ref{sumrul}). The dependence on $\omega_0$ can
be reintroduced later.

As pointed out above, the theoretical expression for the right-hand
side of the sum rule consists of a perturbative part and condensate
contributions: $K=K_{\rm pert}+K_{\rm cond}$. Let us first present the
result for the nonperturbative contributions. The leading terms are
proportional to the quark condensate (dimension $d=3$), the gluon
condensate ($d=4$), and the mixed quark-gluon condensate ($d=5$). For a
consistent calculation at order $\alpha_s$, we calculate the Wilson
coefficients of the quark and gluon condensates to one-loop order, and
the coefficient of the mixed condensate at tree level. The truncation
of the series of power corrections at the mixed condensate seems safe.
The contributions from four-quark operators ($d=6$) are suppressed
relative to the quark condensate by a factor $|\langle\bar q
q\rangle|/T^3\sim 1-5\%$. The calculation is most conveniently
performed using the coordinate gauge $x\cdot A(x)=0$ with the origin
chosen at the position of the velocity-changing heavy quark current. We
find
\begin{equation}
   K_{\rm cond}(T,\infty,y)
   = - {2\alpha_s \langle\bar q q\rangle\,T\over 3\pi}\,
   \Big[ 2 + r(y) \Big] + {\langle\alpha_s GG\rangle\over 96\pi}\,
   \bigg({y-1\over y+1}\bigg)
   - {\langle\bar q\,g_s\sigma_{\alpha\beta} G^{\alpha\beta} q\rangle
   \over 12 T} \,,
\end{equation}
where
\begin{equation}
   r(y) = {1\over\sqrt{y^2-1}}\,\ln\big( y + \sqrt{y^2-1} \big) \,.
\end{equation}

Let us now turn to the perturbative contributions to the sum rule. At
order $\alpha_s$, one has to evaluate the bare quark loop as well as
the seven two-loop diagrams depicted in Fig.~\ref{fig:1}. A new feature
of the present sum rule, as compared to the sum rule for the Isgur-Wise
function considered in Ref.~\cite{Neu3}, is that there is a diagram
($D_7$) where a gluon originates from the covariant derivative
contained in the current. We denote the Borel transformed contributions
of the individual diagrams to the function $K_{\rm pert}(T,\infty,y)$
by $\hat D_i$. Throughout the calculation we use Feynman gauge and
dimensional regularization. The bare quark loop is readily calculated
and gives \cite{Neu2}
\begin{equation}\label{lead}
   \hat D_0 = {2 N_c\,T^D\over(4\pi)^{D/2}}\,
   \bigg({2\over y+1}\bigg)^{D/2}\,\Gamma\big(\case{D}/{2}\big) \,,
\end{equation}
where $D$ is dimension of space-time. The evaluation of the two-loop
corrections is more complicated. For a detailed and systematic
discussion of the techniques used to calculate the two-loop integrals
the reader is referred to Ref.~\cite{Neu3}. The present calculation
proceeds very similar to that in this reference, where the sum rule for
the Isgur-Wise function was derived at two-loop order. It is convenient
to introduce a constant
\begin{equation}
   A = - {4 N_c\,C_F\,g_s^2\over(4\pi)^D}\,
   {(2T)^{2D-6}\over\big[2(y+1)\big]^{D-2}}\,
   \Gamma\big(\case{D}/{2}\big)\,\Gamma\big(\case{D}/{2}-1\big) \,,
\end{equation}
where $C_F=(N_c^2-1)/2 N_c$. Once this quantity is factored out, we
find that the contributions of the first four two-loop diagrams in
Fig.~\ref{fig:1} are the same as the corresponding contributions to the
sum rule for the Isgur-Wise function:
\begin{equation}
   \sum_{i=1}^4 \hat D_i = A\,\bigg\{
   {1\over\epsilon} \bigg[\,{3\over2} - y\,r(y)\bigg]
   + 2\,\Big[ 1 - y\,r(y) \Big]\,\ln\big[2(1+y)\big] + 2 y h(y)
   - 4 + {\cal{O}}(\epsilon) \bigg\} \,.
\end{equation}
We have expanded the result in $\epsilon=(D-4)/2$ and introduced the
function
\begin{equation}
   h(y) = {1\over\sqrt{y^2-1}}\,\Big[ L_2(1-y_-^2) - L_2(1-y_-) \Big]
   + {3\over4} \sqrt{y^2-1}\,r^2(y) \,,
\end{equation}
where $y_-=y-\sqrt{y^2-1}$, and $L_2(x)$ is the dilogarithm. The
calculation of the remaining three diagrams is more cumbersome. It
requires rather elaborate techniques such as Kotikov's method of
differential equations \cite{Koti}. However, remarkable simplifications
take place as we add up the various contributions. In particular all
dilogarithms, which appear in intermediate steps of the calculation,
cancel out. The final result is rather simple:
\begin{equation}
   \sum_{i=5}^7 \hat D_i = A\,\bigg\{ {1\over\epsilon} - 2
   - {2\pi^2\over3} - (y^2-1)\,r^2(y) - (y+1)\,\Big[ 2 + r(y) \Big]
   + {\cal{O}}(\epsilon) \bigg\} \,.
\end{equation}
Except for the last term, this is again the same result as for the
Isgur-Wise function.

The ultraviolet divergent terms in the sum of the seven two-loop
diagrams match with the anomalous dimensions of the heavy-heavy and
heavy-light currents contained in the three-current correlator in
(\ref{correl}). Thus the $1/\epsilon$ pole disappears upon
renormalization of the currents. In the modified minimal subtraction
scheme, the renormalization factors are \cite{Volo,Falk}
\begin{equation}
   Z_{\rm hh} = 1 + {\alpha_s\over 2\pi\hat\epsilon}\,\gamma(y) \,,
   \qquad Z_{\rm hl} = 1 - {\alpha_s\over 2\pi\hat\epsilon} \,,
\end{equation}
where
\begin{equation}
   {1\over\hat\epsilon} = {1\over\epsilon} + \gamma_E
   - \ln{4\pi\over\mu^2} \,,
\end{equation}
and
\begin{equation}
   \gamma(y) = {4\over 3}\,\Big[ y\,r(y) - 1 \Big] \,.
\end{equation}
Our exact two-loop result for the renormalized perturbative part of the
correlator is
\begin{eqnarray}
   K_{\rm pert}(T,\infty,y) &=& Z_{\rm hh}^{-1} Z_{\rm hl}^{-2}\,
    \hat D_0 + \sum_{i=1}^7 \hat D_i \nonumber\\
   &=& {3 T^4\over 8\pi^2}\,\bigg({2\over y+1}\bigg)^2\,\bigg\{
    1 + {\alpha_s\over\pi}\,\bigg[ \Big[ 2 - \gamma(y) \Big]
    \bigg( \ln{\mu\over T} + \gamma_E - {11\over 6} \bigg)
    + {4\pi^2\over 9} + {17\over 3} \nonumber\\
   &&\phantom{ {3 T^4\over 8\pi^2}\,\bigg({2\over y+1}\bigg)^2\,
               \bigg\{ 1 + {\alpha_s\over\pi}\,\bigg[ }
    \mbox{}+ c_{\rm pert}(y) + \delta c(y) \bigg] \bigg\} \,,
\end{eqnarray}
where
\begin{eqnarray}
   c_{\rm pert}(y) &=& {\gamma(y)\over 2} \bigg[ 4\ln 2 - 3
    + \ln{y+1\over 2} \bigg] - {4\over 3} \Big[ y\,h(y)-1 \Big]
    + \ln{y+1\over 2} + {2\over 3}\,(y^2-1)\,r^2(y) \,, \nonumber\\
   \phantom{ \Bigg[ }
   \delta c(y) &=& {2\over 3} + {2\over 3}\,(y+1)\,
    \Big[ 2 + r(y) \Big] - {4\over 9}\,\Big[ y\,r(y) - 1 \Big] \,.
\end{eqnarray}
The function $c_{\rm pert}(y)$ is the same that arises in the
calculation of the sum rule for the Isgur-Wise function.

The final expression for the QCD sum rule (\ref{sumrul}) is obtained
when we reintroduce the continuum threshold $\omega_0$ and write the
result as a dispersion integral. This gives
\begin{eqnarray}\label{xi3sum}
   \bar\Lambda\,\xi_3(y,\mu)\,F^2(\mu)\,e^{-2\bar\Lambda/T}
   &=& {1\over 16\pi^2}\,\bigg({2\over y+1}\bigg)^2
    \int\limits_0^{\omega_0}\!{\rm d}\omega_+\,e^{-\omega_+/T}\,
    \omega_+^3 \nonumber\\
   &\times& \bigg\{ 1 + {\alpha_s\over\pi}\,\bigg[
    \Big[ 2 - \gamma(y) \Big]\,\ln{\mu\over\omega_+} + {4\pi^2\over 9}
    + {17\over 3} + c_{\rm pert}(y) + \delta c(y) \bigg] \bigg\}
    \nonumber\\
   &-& {2\alpha_s \langle\bar q q\rangle\over 3\pi}\,\Big[ 2 + r(y)
    \Big] \int\limits_0^{\omega_0}\!{\rm d}\omega_+\,e^{-\omega_+/T}\,
    + {\langle\alpha_s GG\rangle\over 96\pi}\,
    \bigg({y-1\over y+1}\bigg) \nonumber\\
   &-& {\langle\bar q\,g_s\sigma_{\alpha\beta} G^{\alpha\beta} q
    \rangle\over 12 T} \,.
\end{eqnarray}

It is instructive to compare this to the sum rule for the product
$\bar\Lambda\,\xi(y,\mu)$, which can be obtained from the two-loop
calculation of Ref.~\cite{Neu3}:
\begin{eqnarray}\label{xisum}
   \bar\Lambda\,\xi(y,\mu)\,F^2(\mu)\,e^{-2\bar\Lambda/T}
   &=& {3\over 16\pi^2}\,\bigg({2\over y+1}\bigg)^2
    \int\limits_0^{\omega_0}\!{\rm d}\omega_+\,e^{-\omega_+/T}\,
    \omega_+^3 \nonumber\\
   &\times& \bigg\{ 1 + {\alpha_s\over\pi}\,\bigg[
    \Big[ 2 - \gamma(y) \Big]\,\ln{\mu\over\omega_+} + {4\pi^2\over 9}
    + {17\over 3} + c_{\rm pert}(y) \bigg] \bigg\} \nonumber\\
   &-& {2\alpha_s \langle\bar q q\rangle\over 3\pi}\,\Big[ y\,r(y) - 1
    \Big] \int\limits_0^{\omega_0}\!{\rm d}\omega_+\,e^{-\omega_+/T}\,
    - {\langle\alpha_s GG\rangle\over 96\pi}\,
    \bigg({y-1\over y+1}\bigg) \nonumber\\
   &-& {(2y+1)\over 3}\,{\langle\bar q\,g_s\sigma_{\alpha\beta}
    G^{\alpha\beta} q\rangle\over 4 T} \,.
\end{eqnarray}
Notice that the $\mu$-dependence in (\ref{xi3sum}) and (\ref{xisum}) is
the same. Thus, our explicit calculation is in accordance with the fact
that the function $\eta(y)$ in (\ref{etadef}) is renormalization-group
invariant. We write
\begin{equation}
   \eta(y) = {\xi_3(y,\mu)\over\xi(y,\mu)} = {1\over 3} + \Delta(y) \,,
\end{equation}
and find that $\Delta(y)$ obeys the sum rule
\begin{eqnarray}\label{deltasum}
   \Delta(y)\,\Big[ \bar\Lambda\,\xi(y)\,F^2\,e^{-2\bar\Lambda/T} \Big]
   &=& {\alpha_s T^4\over 12\pi^3}\,\bigg({2\over y+1}\bigg)^2
    \Big[ 11 + 6y + (3+y)\,r(y) \Big]\,
    \delta_3\Big({\omega_0\over T}\Big) \nonumber\\
   &-& {2\alpha_s \langle\bar q q\rangle\,T\over 9\pi}\,
    \Big[ 7 + (3-y)\,r(y) \Big]\,\delta_0\Big({\omega_0\over T}\Big)
    \nonumber\\
   &+& {\langle\alpha_s GG\rangle\over 72\pi}\,
    \bigg({y-1\over y+1}\bigg)
    + {\langle\bar q\,g_s\sigma_{\alpha\beta} G^{\alpha\beta} q\rangle
    \over 18 T}\,(y-1) \,,
\end{eqnarray}
where
\begin{equation}
   \delta_n(x) = {1\over\Gamma(n+1)} \int\limits_0^x\!{\rm d}z\,
   z^n\,e^{-z} \,.
\end{equation}
Note that, since the right-hand side of (\ref{deltasum}) is of order
$\alpha_s$, in this sum rule one is not sensitive to the running of the
quantities $\xi(y)$ and $F$. Their $\mu$-dependence would show up at
order $\alpha_s^2$. For the analysis of the sum rule it is, therefore,
consistent to use
\begin{equation}\label{xilam}
   \bar\Lambda\,\xi(y)\,F^2\,e^{-2\bar\Lambda/T}
   = {9 T^4\over 8\pi^2}\,\bigg({2\over y+1}\bigg)^2\,
   \delta_3\Big({\omega_0\over T}\Big)
   - {(2y+1)\over 3}\,{\langle\bar q\,g_s\sigma_{\alpha\beta}
    G^{\alpha\beta} q\rangle\over 4 T} \,,
\end{equation}
which is obtained from (\ref{xisum}) by neglecting terms of order
$\alpha_s$. Taking the ratio of (\ref{deltasum}) and (\ref{xilam})
reduces to a minimum the systematic uncertainties in the calculation of
$\Delta(y)$.

\section{Numerical Analysis and Conclusions}

In its final form, the sum rule for $\Delta(y)$ very much resembles the
sum rules for the other subleading Isgur-Wise functions $\chi_2(y)$ and
$\chi_3(y)$, which we derived in Ref.~\cite{chi23}. Accordingly, the
numerical analysis proceeds in a similar way. For the QCD parameters we
take the standard values
\begin{eqnarray}
   \langle\bar q q\rangle &=& - (0.23\,{\rm GeV})^3 \,, \nonumber\\
   \langle\alpha_s GG\rangle &=& 0.04\,{\rm GeV^4} \,, \nonumber\\
   \langle\bar q\,g_s\sigma_{\alpha\beta}G^{\alpha\beta} q\rangle
   &=& m_0^2\,\langle\bar q q\rangle ~,~~
    m_0^2 = 0.8\,{\rm GeV^2} \,.
\end{eqnarray}
Our results turn out to be very stable against variations of these
numbers within reasonable limits. Furthermore, we use
$\alpha_s/\pi=0.1$ corresponding to a scale $\mu\approx 2\bar\Lambda
\approx 1$ GeV, which is appropriate for evaluating radiative
corrections in the effective theory. Combining (\ref{deltasum}) and
(\ref{xilam}), we obtain $\Delta(y)$ and hence $\eta(y)$ as functions
of $\omega_0$ and $T$. These input parameters can be determined from
the analysis of a QCD sum rule for the correlator of two heavy-light
currents in the effective theory \cite{Baga,Neu1}. One finds good
stability for $\omega_0=2.0\pm 0.3$ GeV, and the consistency of the
theoretical calculation requires that the Borel parameter be in the
range $0.6<T<1.0$ GeV.

In Fig.~\ref{fig:2}(a) we show the zero recoil value of the form factor
$\eta(y)$ as a function of the Borel parameter, for three different
values of the continuum threshold. For $y=1$ the sum rules simplify
considerably. We find
\begin{equation}
   \eta(1) = {1\over 3} + {14\alpha_s\over 9\pi}\,
   { \displaystyle \delta_3\Big({\omega_0\over T}\Big)
     - {8\pi^2\over 7}\,{\langle\bar q q\rangle\over T^3}\,
     \delta_0\Big({\omega_0\over T}\Big)
   \over \displaystyle \delta_3\Big({\omega_0\over T}\Big)
     - {2\pi^2\over 9}\,{m_0^2\,\langle\bar q q\rangle\over T^5} } \,.
\end{equation}
Neglecting the terms of order $\alpha_s$, we would recover the result
$\eta(1)=1/3$ derived in Ref.~\cite{Neu2}. However, as seen in the
figure these contributions are by no means negligible. They enhance the
form factor by almost a factor 2. It supports the self-consistency of
the sum rule approach that we find stability in essentially the same
region of parameter space that leads to stability of the two-current
sum rules considered in Refs.~\cite{Baga,Neu1}, and of other
three-current sum rules analyzed in Refs.~\cite{chi23,Neu3}.

Over the kinematic range accessible in $\bar B\to D^{(*)}\ell\,\bar\nu$
decays, we show in Fig.~\ref{fig:2}(b) the range of predictions for
$\eta(y)$ obtained for $1.7<\omega_0<2.3$ GeV and $0.6<T<1.2$ GeV. The
numerical analysis confirms our guess that $\eta(y)$ should be a slowly
varying function of order unity, which was the motivation for its
introduction in the first place. In fact, the sum rule predicts that
$\eta(y)\approx 0.6$ essentially independent of $y$. The main
uncertainty comes from the values of $\alpha_s$ and $\omega_0$, which
are not very accurately known. However, one should keep in mind that
there are systematic uncertainties inherent in QCD sum rules which
cannot be estimated by simply varying the input parameters. To be
conservative, we quote our final result as
\begin{equation}\label{etares}
   \eta(y)\approx 0.6\pm 0.2 \,; \qquad 1.0 < y < 1.6 \,.
\end{equation}

This result has important implications for the extraction of
$|\,V_{cb}|$ from $\bar B\to D\,\ell\,\bar\nu$ decays. According to
(\ref{Kfac}), the $1/m_Q$ corrections to the decay rate are
proportional to $[1.06 - 2.18\,\eta(1)]$, and by a fortunate accident
this combination is strongly suppressed for $\eta(1)$ in the range
(\ref{etares}). For the symmetry-breaking corrections to the decay rate
in (\ref{BDrate}), we obtain (we use $\bar\Lambda=0.5$ GeV, $m_c=1.45$
GeV, and $m_b=4.8$ GeV)
\begin{equation}
   1 + S\cdot K\approx 1.00\pm 0.03 \,,
\end{equation}
i.e., at most a few percent. This is comparable to the expected size of
$1/m_Q^2$ corrections \cite{FaNe}. We conclude that the theoretical
uncertainty in the determination of $|\,V_{cb}|$ from this decay mode
is not worse than in $B\to D^*\ell\,\bar\nu$ transitions. Of course,
the experimental measurement of $B\to D\,\ell\,\bar\nu$ near zero
recoil is more difficult. The reward of such a measurement, however,
would be an independent determination of $|\,V_{cb}|$ with surprisingly
small theoretical uncertainties.

Let us finally point out how our sum rule prediction (\ref{etares}) can
be tested experimentally, by a measurement of the form factor ratio
$R_2$ in $\bar B\to D^*\ell\,\bar\nu$ transitions. Using $\eta(y)=0.6$
in (\ref{R2def}) we obtain
\begin{equation}
   R_2\approx 1.0 - 0.2\,\bigg({2\over v\cdot v' + 1}\bigg) \,.
\end{equation}
In Table~\ref{tab:1} we compare this result to the predictions of some
popular quark models, as well as to a recent QCD sum rule calculation
of the weak decay form factors in the full theory. These models give
values for $R_2$ which are substantially larger than ours. In
particular, we note that at $q^2=0$, corresponding to the maximal
velocity transfer, the models give $R_2\ge 1$, whereas we find
$R_2\approx 0.84$. This discrepancy should not be too surprising. Since
we have worked very hard to understand the origin of the
symmetry-breaking corrections, we can hope that our refined sum rule
analysis accounts for such effects in a much more detailed way than the
naive quark models can.

We end this paper with an interesting speculation. Although there is no
reason to believe that it makes any sense to apply the heavy quark
expansion to the $D\to K^*\ell\,\bar\nu$ decay amplitude, we might
still believe in a ``continuity of signs'' and guess that the tendency
$R_2<1$ should persist, and most likely even become more pronounced,
when we imagine changing the heavy quark masses from $m_b$ and $m_c$ to
$m_c$ and $m_s$. This tendency is in fact very consistent with the
experimental value of the form factor ratio obtained from an analysis
of the joint angular distribution in $D\to K^*\ell\,\bar\nu$ decays.
Taking the weighted average of the results reported by the experiments
E691 \cite{Anjo} and E653 \cite{Koda}, we get $R_2^{D K^*}(q^2=0)=0.67
\pm 0.23$. Although we have no right to extrapolate (\ref{R2def}) down
to the strange quark mass, we take this observation as a confirmation
of our prediction that symmetry-breaking corrections suppress
$R_2$.\footnote{The authors of Ref.~\cite{Rosn} apply HQET to $D\to
K^{(*)}\ell\,\bar\nu$ decays and obtain $\eta(1)\approx 0.3\pm 0.4$
(for $\bar\Lambda=0.4$ GeV) from an overall fit to the data.}

In conclusion, we have presented the complete next-to-leading order QCD
sum rule analysis of the subleading Isgur-Wise functions $\xi_3(v\cdot
v',\mu)$ and $\eta(v\cdot v')$, including in particular the two-loop
perturbative corrections. We find that effects of order $\alpha_s$ are
very important and enhance the form factors. Over the kinematic region
accessible in semileptonic decays, the renormalization-group invariant
ratio $\eta(v\cdot v')$ turns out to be essentially constant and equals
$0.6\pm 0.2$. This leads to an almost complete cancellation of the
leading symmetry-breaking corrections to the $\bar B\to
D\,\ell\,\bar\nu$ decay rate at zero recoil, allowing for a reliable
determination of $|\,V_{cb}|$ from this decay mode.

\acknowledgments
M.N. gratefully acknowledges financial support from the BASF
Aktiengesellschaft and from the German National Scholarship Foundation.
Y.N. is an incumbent of the Ruth E. Recu Career Development chair, and
is supported in part by the Israel Commission for Basic Research and by
the Minerva Foundation. This work was also supported by the Department
of Energy, contract DE-AC03-76SF00515.

\begin{table}
   \begin{tabular}{l|c|cccc}
 & our results & ISGW~\cite{ISGW} & BSW~\cite{BSW} & KS~\cite{KS} &
 sum rules~\cite{PBal} \\
\hline
$R_2(1)$ & 0.80 & 0.91 & 0.85 & 1.09 & 0.95 \\
$R_2(y_{\rm max})$ & 0.84 & 1.14 & 1.06 & 1.00 & 1.05
   \end{tabular}
\caption{\label{tab:1}
Predictions for the form factor ratio $R_2$. The zero recoil limit
$y=1$ corresponds to $q^2=(m_B-m_{D^*})^2$, whereas $y_{\rm max}\approx
1.5$ corresponds to $q^2=0$.}
\end{table}

\figure{\label{fig:1}
One- and two-loop perturbative contributions to the sum rule for the
universal form factor $\xi_3(v\cdot v',\mu)$. Heavy quark propagators
are drawn as double lines. The wavy line represents the
velocity-changing heavy quark current $\bar h'\,\Gamma\,i D_\alpha h$.}

\figure{\label{fig:2}
Numerical evaluation of the sum rule (\ref{deltasum}): (a) dependence
of the zero recoil form factor $\eta(1)$ on the Borel parameter for
different values of the continuum threshold; (b) the function $\eta(y)$
for $0.6<T<1.2$ GeV and $1.7<\omega_0<2.3$ GeV.}

\end{document}